\documentclass[fleqn,usenatbib]{mnras}
\usepackage{multicol}
\usepackage[T1]{fontenc}
\usepackage{ae,aecompl}
\usepackage{graphicx}	
\usepackage{amsmath}	
\usepackage{amssymb}	
\usepackage{multirow}
\title[Investigation of four RRATs at the frequency of 111 MHz]{Investigation of four rotating radio transients properties at 111 MHz}
\author[S. A. Tyul'bashev et al.]{
S. A. Tyul'bashev$^{1}$\thanks{E-mail: serg@prao.ru}
T. V. Smirnova$^{1}$
E.A.Brylyakova$^{1,2}$
and M.A. Kitaeva$^{1}$
\\ 
$^{1}$ P.N. Lebedev Physical Institute of the Russian Academy of Sciences, Astro Space Center, Pushchino Radio Astronomy Observatory\\
Radiotelescopnaya 1, Moscow reg., Pushchino, 142290, Russia \\
$^{2}$ P.N. Lebedev Physical Institute of the Russian Academy of Sciences\\
Leninskii pr. 53, Moscow, 119991  Russia \\
}

\date{Accepted XXX. Received YYY; in original form ZZZ}
\pubyear{2021}
\begin{document}
\label{firstpage}
\pagerange{\pageref{firstpage}--\pageref{lastpage}}
\maketitle 
\begin{abstract} 
The analysis of individual pulses of four rotating radio transients (RRATs), previously discovered in a monitoring survey running for 5.5 years at the frequency of 111 MHz, is presented. At a time interval equivalent to five days of continuous observations for each RRAT, 90, 389, 206, and 157 pulses were detected in J0640+07, J1005+30, J1132+25, and J1336+33, respectively. The investigated RRATs have a different distribution of the pulses amplitude. For J0640+07 and J1132+25, the distribution is described by a single exponent over the entire range of flux densities. For J1005+30 and J1336+33, it is  a lognormal function with a power law tail. For J0640+07 and J1005+30, we have detected pulses with a signal-to-noise (S/N) ratio of few hundreds. For J1132+25 and J1336+33, the S/N of the strongest pulses reaches several tens. These RRATs show strong changing of character of emission. When strengths of pulse amplitudes significantly changed, we see long intervals of absence of emission or its strong attenuation. The analysis carried out in this work shows that it is possible that all the studied RRATs are, apparently, pulsars with giant pulses.
\end{abstract}

\begin{keywords}
	pulsars: general
\end{keywords}

\section{Introduction}

Rotating radio transients (RRATs) were discovered in 2006 (\citeauthor{McLaughlin2006}, \citeyear{McLaughlin2006}) as sources of sporadic bursts of dispersed pulses followed by no detectable emission for many rotations (sometimes minutes
to hours) (Keane, 2016; Bhattacharyya et al., 2018).  They are Galactic neutron stars with extreme emission variability. Single pulse rates are in the range of
a few pulses to a few hundred pulses per hour. The nulling fraction of RRATs can exceed 99 percent.
 The average magnetic fields of RRATs and the average periods are higher than those of ordinary second pulsars (\citeauthor{McLaughlin2009}, \citeyear{McLaughlin2009}, \citeauthor{Cui2017}, \citeyear{Cui2017}). Other qualities of RRATs are similar to those of pulsars with similar periods. According to the paper  \citeauthor{Burke-Spolaor2010} (\citeyear{Burke-Spolaor2010}) Galactic z-distribution and pulse width distributions are the same as for ordinary pulsars.   

Long-term studies of RRATs are still limited (\citeauthor{Palliyaguru2011}, \citeyear{Palliyaguru2011}, \citeauthor{Bhattacharyya2018}, \citeyear{Bhattacharyya2018}, \citeauthor{Mickaliger2018}, \citeyear{Mickaliger2018}, \citeauthor{Shapiro-Albert2018}, \citeyear{Shapiro-Albert2018}, \citeauthor{Brylyakova2020}, \citeyear{Brylyakova2020}). The authors of the paper  \citeauthor{Palliyaguru2011} (\citeyear{Palliyaguru2011}), \citeauthor{Cui2017} (\citeyear{Cui2017}) analyzed the short and long-term variability of eight RRATs studied over a 5.5-year interval at frequency 1400 MHz. The paper shows that there are periodicities from 30 minutes to years in pulse arrival time for six investigated RRATs and RRAT pulses appear randomly at small time intervals. 

In the paper \citeauthor{Cui2017} (\citeyear{Cui2017}) for eight RRATs, a series of observations from 1.5 to 15 years were used to obtain accurate estimates of the rotational parameters, as well as to obtain histograms of the pulses amplitude distributions. It turned out that the typical pulse distribution is lognormal, as for ordinary pulsars. A power law tail in the pulses energy distributions typical for pulsars with giant pulses have observed for several RRATs. Results of research on the pulses distribution by energy in the paper \citeauthor{Mickaliger2018} (\citeyear{Mickaliger2018}) are the same as in \citeauthor{Cui2017} (\citeyear{Cui2017}), but for 14 RRATs. 
A study of three RRATs over an eleven-year series of observations was presented in  \citeauthor{Shapiro-Albert2018}, (\citeyear{Shapiro-Albert2018}). It shows that the energy distribution for two RRATs is lognormal, and for one, it is lognormal, and an additional a power component is observed. 

Using simultaneous observation of single pulses of RRAT J2325-0530 at Murchison Widefield Array (MWA, 154 MHz) and Parkes radio telescopes (1.4 GHz)  \citeauthor{Meyers2019} (\citeyear{Meyers2019}) measure the spectral index $\alpha = 2.2\pm 0.1$. \citeauthor{Shapiro-Albert2018} (\citeyear{Shapiro-Albert2018}) provided single-pulse based spectral index ($\alpha$) for three studied RRATs which are between $0.6$ and $1.2$. \citeauthor{Taylor2016} (\citeyear{Taylor2016}) measure $\alpha$ about $0.7$ across band of LWA1 (35 - 80 MHz) for RRAT J2325-0530 which is substantially shallow than in \citeauthor{Meyers2019} (\citeyear{Meyers2019}). It is possible that a flattering may occur at f < 150 MHz. 

The eight-year observations for three RRATs are considered in  \citeauthor{Bhattacharyya2018} (\citeyear{Bhattacharyya2018}). It is shown in this paper that there may be long-term trends of changes in the observed arrival rate of pulses. The average number of observed pulses varies 1.5-2 times both in the direction of decreasing and increasing of the number of observed pulses over the full observation interval. It is noted in this paper that in the absence of strong pulses, one from investigated  RRATs shows weak periodic emission. 

The five-year observation interval at the frequency 111 MHz have used to study the RRAT J0139+33 (\citeauthor{Brylyakova2020}, \citeyear{Brylyakova2020}). We have shown that the energy distribution of pulses in the RRATs is described by bimodal (broken) power law, which is typical for some pulsars with giant pulses (\citeauthor{Smirnova2012}, \citeyear{Smirnova2012}, \citeauthor{Kazantsev2017}, \citeyear{Kazantsev2017}).

Summarizing the above mentioned observations, we can come to conclusion that the number of observed pulses can significantly change for RRAT at certain time intervals, trends can be observed that show an increase or decrease in the number of observed pulses over the entire observation interval. The pulse energy distribution function is usually lognormal, although a power law tail is sometimes observed. There is also one case of a broken power law distribution.

In 2018, when processing semi-annual daily observations obtained at the Large Phased Array (LPA) radio telescope of the Lebedev Physical Institute (LPI), 33 RRATs were detected by their individual dispersed pulses (\citeauthor{Tyulbashev2018}, \citeyear{Tyulbashev2018}, \citeauthor{Tyulbashev2018a}, \citeyear{Tyulbashev2018a}). Out of this sample, we have selected for further studies four RRATs that have long nullings or strong individual pulses. These RRATs were discovered in the monitoring survey carried out at the LPA LPI at a wavelength of 2.7 meters. They have not yet been confirmed by observations on other instruments. In particular, they were not detected in the search for pulsars on LOFAR (\citeauthor{Sanidas2019}, \citeyear{Sanidas2019}) at the wave length 2.2 m. 

\section{Observations}

Round-the-clock daily monitoring observations were carried out on the upgraded antenna of the LPA LPI since August 2014. LPA LPI consists from 16384 wave dipoles. These are 256 lines each of which has 64 dipoles. We have used Butler matrices to form a beams in the Meridian plane. Details about the modernization of the LPA LPI are given by  \citeauthor{Shishov2016} (\citeyear{Shishov2016})  and \citeauthor{Tyulbashev2016} (\citeyear{Tyulbashev2016}).

The main purpose of the monitoring program is daily observations under the 'Space Weather' program. In this program, daily observations of about 5000 compact (scintillating) radio sources are carried out (\citeauthor{Shishov2016}, \citeyear{Shishov2016}). Observations made for the 'Space Weather' program can also be used to study pulse emission (\citeauthor{Brylyakova2020}, \citeyear{Brylyakova2020}).

The antenna operates at a central frequency of 110.3 MHz. The full band 2.5 MHz is divided into 32 frequency channels with a channel width of 78 kHz. The sampling interval is equal to 12.5 ms. Observations are done  simultaneously in 96 antenna beams covering declinations from $-9\degr$ to $+42\degr$. The data obtained are used to search for new RRATs and pulsars. 

Digital recorders were created for these observations do not allow us to get more frequency channels and readout time less then 12.5 ms. The accuracy of the timestamps are determined by the quartz oscillator. Therefore raw data are not optimal for pulsar and RRAT investigations.

Six times a day a signal of known temperature (calibration signal) is applied to the antenna input in the form of "OFF-ON-OFF", this allows to equalize the gain between frequency channels (\citeauthor{Tyulbashev2020}, \citeyear{Tyulbashev2020}). As the data analysis showed, during a 2-hour observation, the change in the calibration signal amplitute does not exceed $5\%$. Since the closest calibration signal to the source is used for session calibration, the detected pulse amplitudes have errors of less than $5\%$. The width of beam directivity pattern of the LPA LPI at half power is equal to $3.5 minutes/ \cos(\delta)$ ($\delta$ is a source declination). 

The typical sensitivity of the LPA LPI telescope when observing ordinary pulsars in a single observation session is 6-8 mJy, if the pulsar is outside the plane of the Galaxy and 15-20 mJy, if the pulsar is in the plane of the Galaxy  (\citeauthor{Tyulbashev2016}, \citeyear{Tyulbashev2016}). The detection limit for single pulses with the selected sampling interval 12.5 ms is 2.1 Jy at $S/N$ (signal to noise ratio) $= 7$ (\citeauthor{Tyulbashev2018}, \citeyear{Tyulbashev2018}). For 5.5 years of continuous monitoring, the equivalent continuous observation time is approximately five days for each point in the sky that falls within the observation area. 

Additional information about observation modes with LPA LPI, on implementation of independent radio telescopes based on a single antenna field and other technical information about the capabilities of the LPA LPI after its modernization can be found in the papers \citeauthor{Shishov2016} (\citeyear{Shishov2016}),  \citeauthor{Tyulbashev2016} (\citeyear{Tyulbashev2016}).

\section{Data analysis} 

LPA LPI have a lot of specifics as antenna array. Therefore, special software were created to process the observations, taking into account these features. We have hosted this software on Github\footnote{https://github.com/Elinxt/rrat$\_$pulsars}. 

We followed by standard way for search of RRAT pulses in the obtained data. After equalizing the amplification between the frequency channels, we incoherently dedispersed the data to the known dispersion measure ($DM$) of the studied RRATs, subtracted the baseline and removed radio frequency interference (RFI), and detected single pulse from the frequency averaged time series.  Details of RFI removal are described in \citeauthor{Tyulbashev2020} (\citeyear{Tyulbashev2020}). 

To get the baseline, a recording section containing a pulse and having a duration of four seconds was taken. This time interval is much longer than the duration of the RRAT pulse, which makes it possible to adequately determine the baseline (\citeauthor{Brylyakova2020}, \citeyear{Brylyakova2020}). First, the frequency channels were summed up without shifts, that is, assuming a $DM$ is equal zero. The resulting points were fit by a polynomial. Then the frequency channels were summed up with shifts corresponding to the $DM$ of the RRAT. The resulting polynomial was subtracted from the resulting series of points. The root-mean-square deviation ($\sigma_n$) of the noises was determined by the area outside the pulse and the S/N was taken as the value of the peak pulse amplitude divided by the estimate of $\sigma_n$ (see details in the paper  \citeauthor{Brylyakova2020} (\citeyear{Brylyakova2020}). For J1005+30, J1132+25 and J1336+33, the pulse detection threshold was set at $S/N$ of 6. For J0640+07 the threshold was increased to $S/N \ge 7$ since a lower threshold resulted in an overwhelming number of false positive detections. 

For further work, pulses were selected that were located at an interval of $\pm 1.5$ minutes from the pulsar passing through the center of the beam. For each pulse, corrections to the detected were made related to the features of LPA LPI, as an antenna array (\citeauthor{Tyulbashev2016}, \citeyear{Tyulbashev2016}, \citeauthor{Shishov2016}, \citeyear{Shishov2016}). These corrections attempt to mitigate the fact that the direction of the formed radiation pattern does not coincide with the direction to the source (details see in \citeauthor{Brylyakova2020} (\citeyear{Brylyakova2020})). For all detected signals, the average profile and dynamic spectrum files were saved for each observation session. The final selection of pulses was done manually. The average profile was determined by summing up a signal with a given period, if it is known. 

As a result of processing, data series were obtained in which the time of arrival of the RRAT pulse and its amplitude ($I_k$), normalized to the $\sigma_n$ were recorded for each date over the entire observation interval. These series were analyzed to study time variations, the number of pulses, and their amplitudes.  Since the data is calibrated based on the calibration signal, it reflects the correct distribution of amplitudes over a long time interval.

\section{Results} 
Figs 1 and 2 show the values of the amplitudes in units of $S/N$, and the number of recorded pulses per session for the selected objects, depending on the sequence of days starting from the first day indicated in Table 1.

\begin{figure*}
	\includegraphics[
	height= 0.4\textheight]{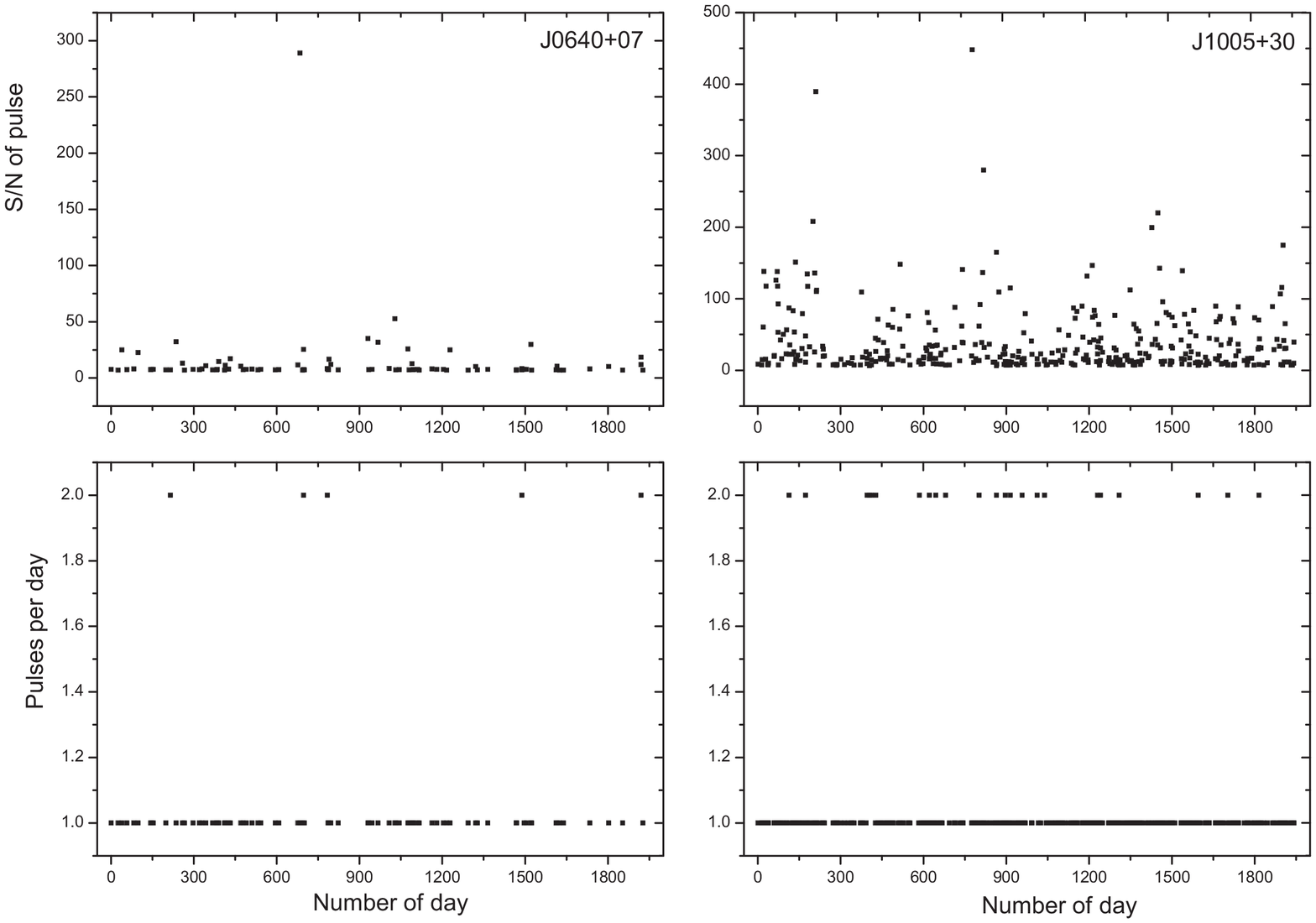}
	\caption{Vertical axis is the pulse amplitude in $S/N$ units (top) and the number of registered pulses (bottom). 
    Horizontal axis is defined as the number of the corresponding days. The first day with number 0 corresponds to JD = 2456920 for J0640+07 and JD = 2456898 for J1005+30.}
\end{figure*} 

\begin{figure*}
	\includegraphics[
	height= 0.4\textheight]{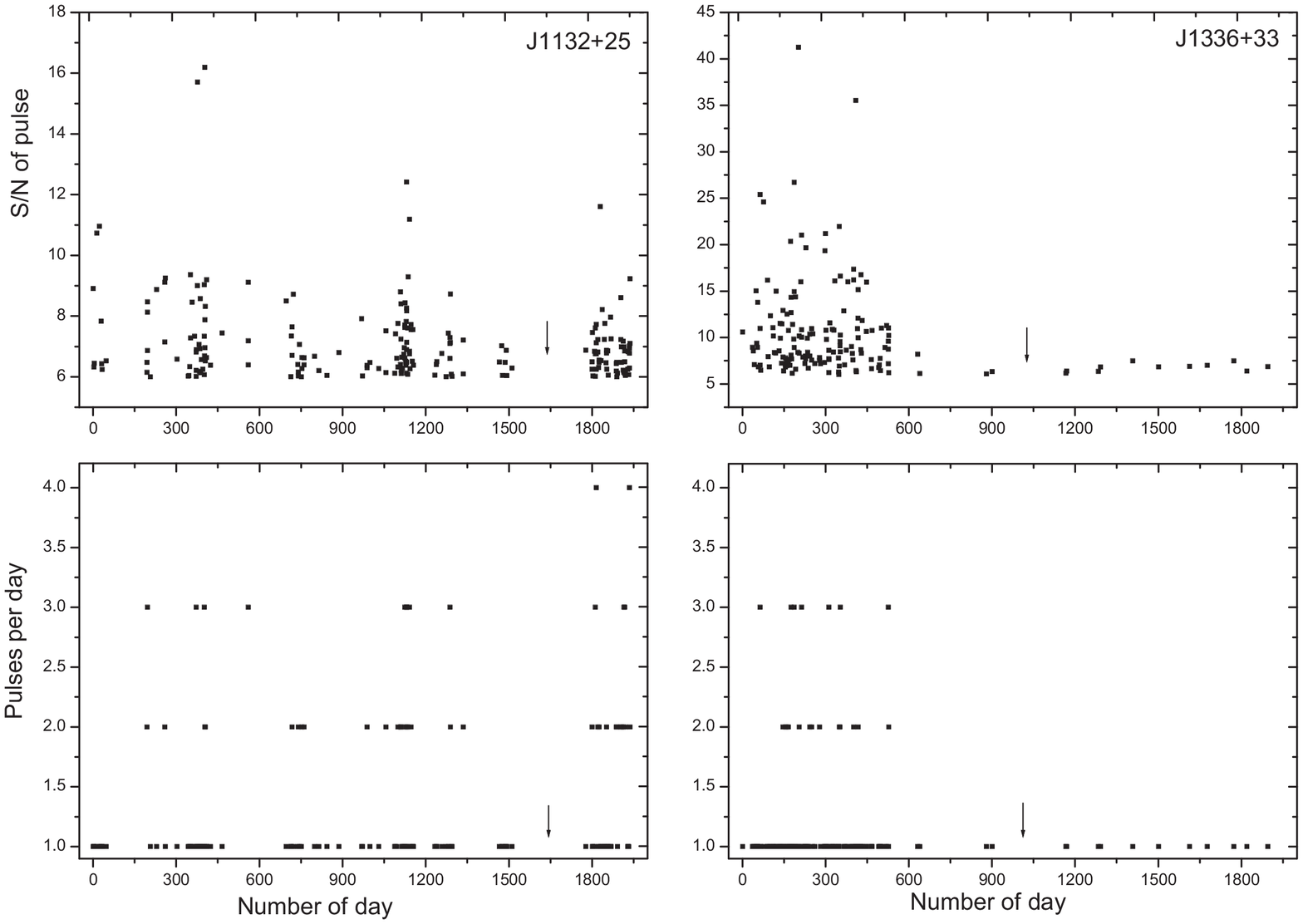}
	\caption{Horizontal and vertical axes are similar to the axes in Fig.1. The first day with number 0 corresponds to JD = 2456896 for J1132+25 and for J1336+33.}
\end{figure*} 

\begin{figure*}
	\includegraphics[
	height= 0.4\textheight]{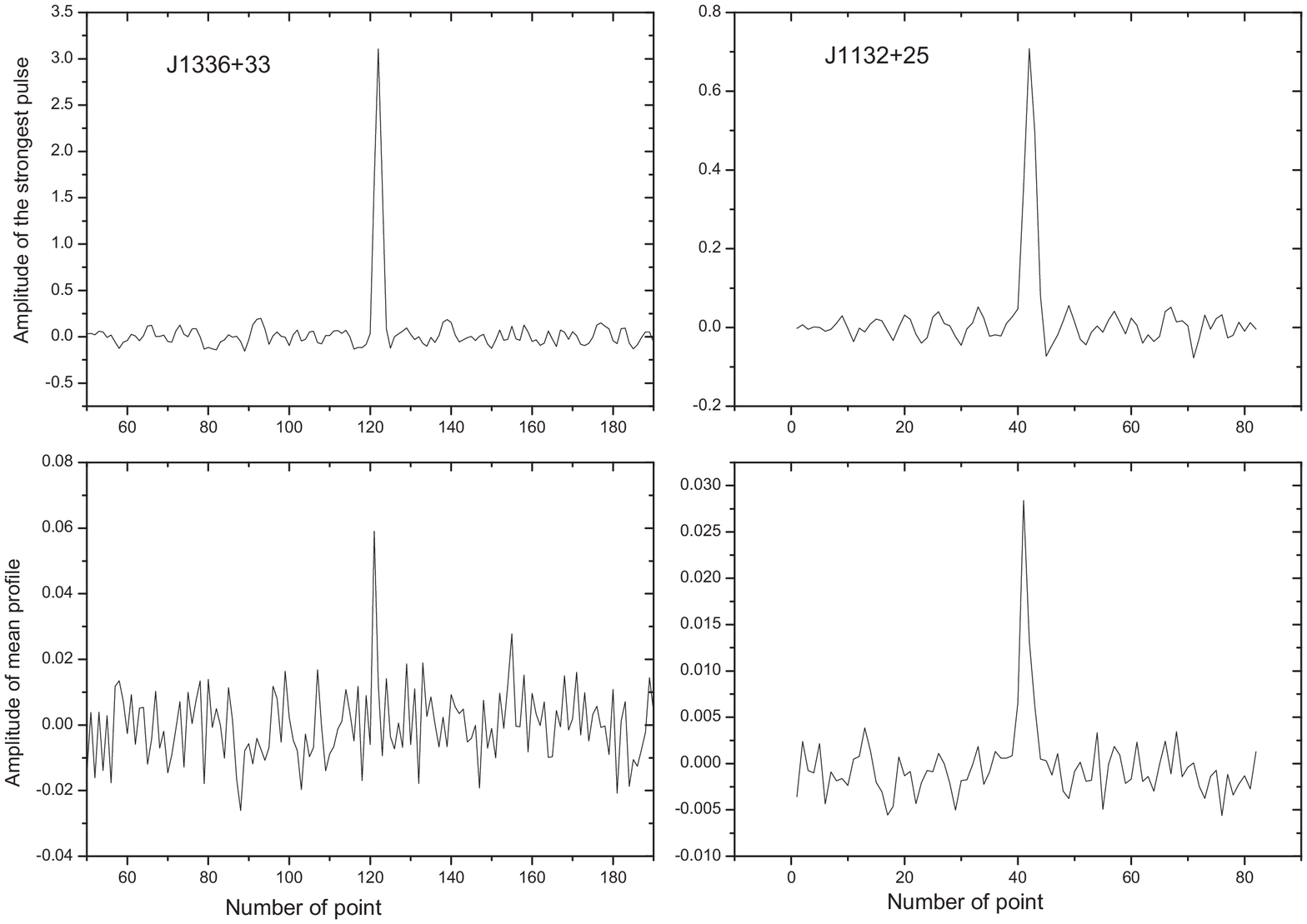}
	\caption{Vertical axis (amplitude) is shown in arbitrary units. Horizontal axis is a time in points ($\delta t = 12.5$ ms). The strongest pulses J1132+25 (top, right) and J1336+33 (top, left) and the average profiles of these RRATs (bottom).}
\end{figure*}  

\begin{figure*}
	\includegraphics[
	height= 0.2\textheight]{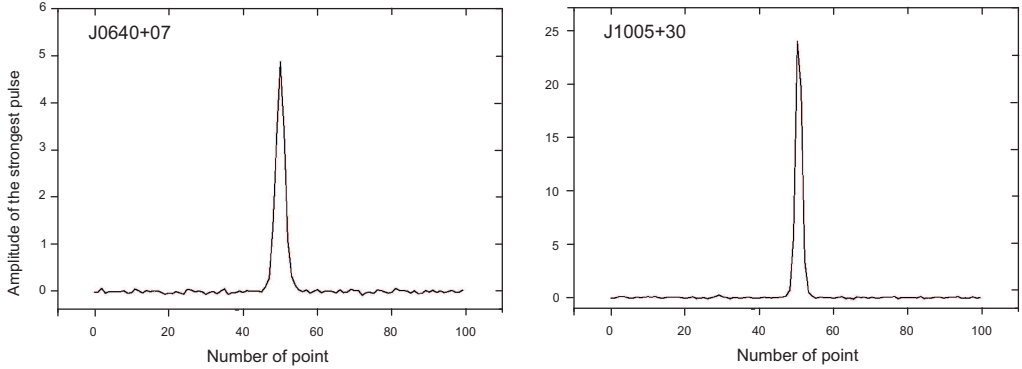}
	\caption{Vertical axis (amplitude) is shown in arbitrary units. Horizontal axis is a time in points ($\delta t = 12.5$ ms). The strongest pulses for J0640+07 (left) and J1005+30 (right).}
\end{figure*}  

\begin{figure*}
	\includegraphics[
	height= 0.4\textheight]{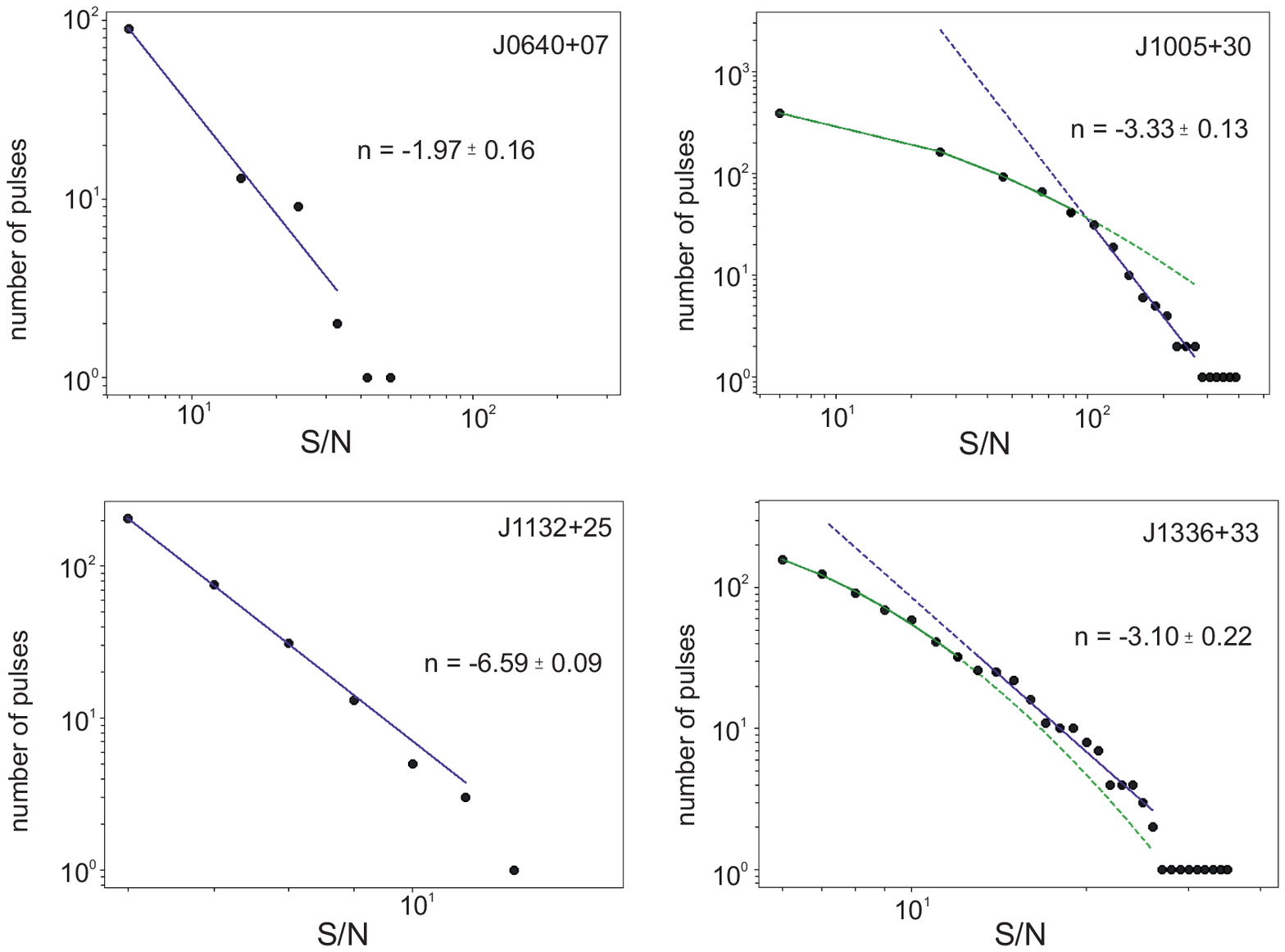}
	\caption{Integral distribution functions of the number of pulses (vertical axis) vs. the pulse amplitude in $S/N$ units (horizontal axis) in a logarithmic scale. The data points are approximated by different laws (see Table 2) using the least squares method.  Obtained power coefficients $n$ with their errors are indicated in the figure.}
\end{figure*}

Table 1 shows information about the observed pulses. In the first column, there is the name of the RRAT. Columns 2-3 show the period ($P_1$) and the dispersion measure of the RRAT, and columns 4-8 show the date when the first pulse was detected, total number of observation days between the first and last recorded pulse ($N_{all}$), number of detected pulses ($N_p$), the ratio of the number of days in which the pulses were detected to the number of all days of observation, i.e. intermittency factor ($f_{int}$), maximum continuous duration absence of pulses in days ($N_{max}$).  Columns 9-12 show average values of amplitudes (in $S/N$ units) for all detected pulses exceeding the specified threshold, $\langle I \rangle_1$, average values taking into account intermittency factor, $\langle I \rangle_2 = \langle I \rangle_1 \times f_{int}$, maximum value of S/N for the entire observation period, $I_{max}$, modulation index  
$m = \sum\limits_{i=1}^N [(I_i - \langle I \rangle_1)^2 /(N - 1)]^{1/2}/ \langle I \rangle_1 $, where $N$ is the number of pulses, and $I_i$ is the amplitude of pulse $i$.

\begin{table*}
	\centering
	\caption{Summary of recorded pulses data.}
	\label{tab:example_table}
	\begin{tabular}{cccccccccccc}
			\hline
PSR & $P_1$ (s) & $DM$ (pc/cm$^3$) & $First$ $date$ & $N_{all}$ (days) & $N_p$ & $f_{int}$  & $N_{max}$ (days) & $\langle I \rangle_1$ & $\langle I \rangle_2$ & $I_{max}$ & m\\
	\hline
J0640+07 & - & 52 & 19 Sept. 2014 & 1927 & 90 & 0.044 & 107 & 14 & 0.62 & 289 & 2.2\\
J1005+30 & - & 17.5 & 24 Sept. 2014 & 1942 & 389 & 0.188 & 31 & 38.3 & 7.2 & 448 & 1.25\\
J1132+25 & 1.002 & 23 & 26 Aug. 2014 & 1937 & 206 & 0.075 & 265 & 7.13 & 0.53 & 16.2 & 0.2\\
J1336+33 & 3.013 & 8.5 & 26 Aug.2014 & 1896 & 157 & 0.069 & 256 & 10.4 & 0.71 & 41.3 & 0.5\\
\hline
\end{tabular}
\end{table*}

The sources in Table 1 can be divided into two samples. In the first sample, there are sources with high modulation and relatively short intervals when no pulses were observed, in the second sample, there are sources with low modulation and intervals when pulses are not detected for many months.

\subsection{J0640+07 and J1005+30 -- RRATs with giant pulses?}

Number of known pulsars with giant pulses (GP) very small compared to the nearly three thousand pulsars published in the ATNF\footnote{https://www.atnf.csiro.au/people/pulsar/psrcat/} catalog. According to Table 1 (\citeauthor{Kazantsev2018}, \citeyear{Kazantsev2018}) with citations of original discoveries of GP  the number of such pulsars were sixteen. The original papers consider a number of features that distinguish pulsars with giant pulses from ordinary pulsars  \citeauthor{Sutton1971} (\citeyear{Sutton1971}), \citeauthor{Kinkhabwala2000}, (\citeyear{Kinkhabwala2000}), \citeauthor{Soglasnov2004}, (\citeyear{Soglasnov2004}), \citeauthor{Hankins2007}, (\citeyear{Hankins2007}), \citeauthor{Kazantsev2018} (\citeyear{Kazantsev2018}). If we consider the radio range, such features may be (\citeauthor{Kazantsev2018} (\citeyear{Kazantsev2018})): high peak flux densities in the pulse compared to the peak flux density accumulated in the average profile (the ratio more  than thirty); the power law distribution of pulse energy; the narrowness of the giant pulse compared to the average profile, sometimes it is an extreme narrowness of the pulse; high degree of polarization; extremely high brightness temperatures for the narrowest (nanoseconds) pulses. However, not all of these features are simultaneously observed in every pulsar with giant pulses. The issue of the ratio of the amplitudes (or energies) of pulses and the average profile for determining the "gigantism" of pulses is also not strictly defined. So in the paper of  \citeauthor{Karuppusamy2012} (\citeyear{Karuppusamy2012}), it is considered that for giant pulses, their energy must exceed the energy of the average pulse by more than 10 times.

From the set of listed "gigantism" features in our monitoring observations, we can check the distribution of pulses by amplitudes or by energies and determine the peak flux densities. It is also necessary to obtain the average profile of the pulsar and compare its amplitude with the amplitude of the strongest observed pulses. If the ratio of amplitudes is at least 30 and the ratio of energies is more than 10, then the studied pulsar is a candidate for pulsars with giant pulses. For RRATs J0640+07 and J1005+30, the period has not yet been determined. The average profile with the lack of timing and a small number of observed pulses per session (one or two) could not be obtained for either J0640+07 or J1005+30 and that's why the ratio of amplitudes was determined from the detected pulses as $I_{max} / \langle I \rangle_2$. As will be shown below, for J1132+25, J1336+33 the value $\langle I \rangle_2$, corresponds well enough to the amplitude of the average profile. This gives the reason to believe that also for the RRATs considered here, obtained value of $\langle I \rangle_2$ determines the amplitude of the average profile. As follows from Table 1, the ratio $I_{max} / \langle I \rangle_2$ is equal to 466 (J0640+07) and 62 (J1005+30) for the strongest pulses. Fig. 1 shows dependence of the signal amplitude in $S/N$ units (top) and the number of registered pulses (bottom) vs. time in days. These amplitude ratios and their distribution functions (analysed in Section 4.3) are GP-like features.

The RRATs J0640+07 and J1005+30 are characterized by a high modulation index associated with strong deviations of the amplitude from the mean value. The emission of J0640+07 and J1005+30 is sporadic. For J0640+07, before the pulse with the maximum amplitude $S/N = 289$ (Fig.1, on the left), eight days before it and another 9 days after it not a single pulse was registered. In this session, it was also the only one. For J1005+30, on the days of the largest amplitudes (Fig. 1, on the right) no pulses above the threshold were registered for several days before it and after it. Figure 3 shows the strongest pulses for J0640+07 and J1005+30. The average rate of occurrence of pulses per observational session is 0.05 (J0640+07) and 0.20 (J1005+30). Average number of pulses during three minutes session of observation $N_p/(f_{int} \times N_{all})$ is equal to 1.05 and 1.06 if we registered pulses during sessions (Fig. 1, bottom).  

\subsection{J1132+25 and J1336+33 -- \textbf{highly intermittent RRATs} }

RRATs J1132+25 and J1336+33 differ from RRATs J0640+07 and J1005+30. They have a sharp change in amplitudes, accompanied by the absence of emission or its sharp weakening for a long time like intermittent pulsars. For RRAT J1132+25 there is long continuous time intervals (of 265 sessions, marked by arrows in Fig. 2), when there are no pulses exceeding the specified threshold. The same figure also shows the highly sporadic nature of these RRATs emission, when long periods of quiescence are replaced by a significant increase in both the amplitude and the number of observed pulses. It turned out that, as for J0640+07 and J1005+30, they have features of giant pulses: $I_{max} / \langle I \rangle_2 = 30.4$ for J1132+25 and $I_{max} / \langle I \rangle_2 = 57.9$ for J1336+33.

In contrast to RRATs J0640+07 and J1005+30, for which we were not able to obtain average profiles, for J1132+25, J1336+33, the usual pulsar type periodic emission was detected previously (the periods shown in Table 1 obtained in the paper \citeauthor{Tyulbashev2018a} (\citeyear{Tyulbashev2018a})). Their average profiles and the strongest pulses are shown in Fig. 4. When obtaining average profiles, at least twenty sessions were randomly selected for each source, during which no strong pulses were observed to exclude influence of them on the profile. For these days, average profiles were obtained after averaging the signal with a known period. It is indicated that weak emission exist even we do not see strong pulses. Since the individual strong pulses and the resulting average profiles were normalized by the calibration signal, it is possible to determine the ratio of the amplitude in the pulse to the amplitude of the average profile.

As can be seen from Fig. 4, the width of the individual pulse and of the average profile at the FWHM level for J1132+25 and J1336+33 are comparable, and the pulse shapes are similar, so we can assume that the ratio of their amplitudes corresponds to the ratio of their energies. We calculated energies as the integral over the pulse or average profile up to the 6 sigma noise level.
Using the obtained shapes of the profile and of individual pulse, we obtained the ratio of the energy in the pulse to the energy of the average profile: $E = E_{pulse} / E_{profile}$. 
These values for the strongest pulses are equal to: $E = 25.3$ (J1132+25), $E = 50.1$ (J1336+33). 
We note that these energies are close (the difference is of the order of $10\%$) to the quantity $I_{max} / \langle I \rangle_2$, and so we can assume that $\langle I \rangle_2$ approximates the amplitude of the average profile.
These ratios of the energy in the pulse to the energy of the average profile and also the amplitude distribution analysed in Section 4.3 tell us that RRATs could be GP emitters.

The average profiles of pulsars J1132+25 and J1336+33, obtained at a three-minute interval and then averaged over several sessions, perhaps do not reflect the real average profile, which should be determined on at least half an hour interval. In addition, the average profile we use is affected by polarization, since the antenna receives linearly polarized emission, and if the period of the Faraday rotation of the polarization plane is comparable to or longer than the receiver band, then the amplitude of such a profile can vary by several times from session to session. Such behavior shows PSR B0950+08 (\citeauthor{Smirnova2012}, \citeyear{Smirnova2012}) observed  at 112 MHz. This pulsar has about 70 \% linear polarization of emission and 15 MHz period of the Faradey rotation which considerably exceeds the receiver bandwidth. This pulsar has one of the smallest rotation measure (RM = 1.35$rad/m^2$) and so the largest profile amplitude instability from session to session compare to other pulsars. Polarization study of 22 known RRATs by \citeauthor{Caleb2019} (\citeyear{Caleb2019}) with the Parkes telescope had shown that average linear polarization fraction
of 40 per cent. Individual single pulses were observed to be up to 100 per cent linearly polarized. At our low frequency the level of poliarization can be about this value.  However the value $\langle I \rangle_2$ obtained from many pulses will correctly reflect the amplitude of the average profile. 

An influence of diffraction scintillations on the profile is very small in our case, since the de-correlation bandwidth for the selected RRATs, according to our estimates (less than a few kHz), is significantly narrower than the receiver band (2.5 MHz). Refractive scintillations for selected objects at our frequency have a scale of the order of a year or longer, so they also do not affect the value of the average profile amplitude. Our estimation of the diffractive scintillation bandwidths $f_{dif}$ and the time scale of refractive scintillation is based on the YMW16 Galactic electron density distribution model (\citeauthor{Yao2017}, \citeyear{Yao2017}). Our measurements are in agreement with previous low-frequency estimates from low-DM pulsars  (\citeauthor{Malofeev1995}, \citeyear{Malofeev1995}). 

For these two sources, as well as for J0640+07 and J1005+30, we have a small value of the ratio of the number of days in which pulses were detected to the total number of all observation days:  $f_{int}$ = 0.075 and 0.069, as well as a small number of pulses exceeding the threshold in the observation session. The average rate of occurrence of pulses per minute is 0.037 (J1132+25) and 0.027 (J1336+33). Average number of pulses per minute is equal to 0.47 (J1132+25) and 0.4 (J1336+33) if we registered pulses during sessions  (Fig.2, bottom). 

The modulation index for J1132+25 and J1336+33 is smaller than 1. As can be seen from Fig. 2, the emission has the flashy character, when the signal amplitude changes dramatically. Especially strong is the change in amplitude for J1336+33, when, after 524 days of high activity, only weak and single pulses are recorded with large intervals of their complete absence (265 and 256 days) up to the last day of observations.

\subsection{Distribution function of pulses}

The number of detected pulses in the studied RRATs is low. Early observations of four RRATs (\citeauthor{McLaughlin2006}, \citeyear{McLaughlin2006}) with a small number of detected pulses (from 11 to 229) showed a power-law distribution of the pulses over the amplitudes. In the paper \citeauthor{Mickaliger2018} (\citeyear{Mickaliger2018}), the dependencies of the same RRATs are shown by the sum of two lognormal distributions.

We checked different fits in the resulting histograms: power law, broken power law, lognormal distribution, lognormal distribution with a power tail. 

$$ f(x)= q/x \cdot e^{-(ln(x)-w)^2/2c^2}$$

$$f(x)=a\cdot x^b,$$

where q, w, c, a, b are fitting parameters of lognormal and power low distributions.

For the fitting, we used the Levenberg-Marquardt non-linear least squares method (see library of Python: LMFIT). Table 2 contains results of model testing using  the Akaike information criterion (AIC\footnote{https://github.com/lmfit/lmfit-py/}) corrected for small sample size (AICc, see details in \citeauthor{Burnham2002} (\citeyear{Burnham2002}). Bold text in the table indicates the best fit, which corresponds to the lowest number in the Table 2 for different models. 

\begin{table*}
	\centering
	\caption{Tests of different laws of pulse amplitude distribution. The values in the Table correspond to the AICc score for each model per pulsar.}
	\label{tab:distribution_of_pulses}
	\begin{tabular}{ccccc}
			\hline
RRAT & power  & broken     & lognormal + & lognormal \\
name &        & power      & power tail  &   \\
	\hline
J0640+07 & {\bf 9}  & 36 & 39 & 53 \\
J1005+30 & 85 & 62 & {\bf 16} & 49 \\
J1132+25 & {\bf 5} & 14 & 30 & 21 \\
J1336+33 & 65  & 51 & {\bf 31} & 37 \\
\hline
\end{tabular}
\end{table*}

For RRAT J0640+07, the number of detected pulses is less than one hundred. Therefore, there are no pulses in the majority of the defined S/N bins.  We have increased the size of the bin by nine times, to build a smoothed distribution.

In Fig.~5 integral distribution functions are given for all four sources in a logarithmic scale. It is clear that J0640+07 and J1132+25 have a power law spectrum. J1005+30 (up to $S/N = 72$) and J1336+33 (up to $S/N = 12$) have a lognormal distribution and the power law distribution above these S/N.

Power or power tail dependencies for distribution functions are additional indirect signs of pulsars with giant pulses. However, the small number of detected pulses does not give us complete confidence to say that investigated RRATs are a pulsar with giant pulses.

\section{Discussion}

All four RRATs have a number of common features: 1) flashy character when the amplitude of the signal increases by several times compared with adjacent sessions (for J1132+25) and even 1-2 orders of magnitude (for J0640+07, J1005+30, J1336+33); 2) a large number of days when there are no pulses above the specified detection threshold (long quiescence periods): $f_{int}$ < 0.2; 3) a significant predominance of sessions with only one pulse for 3 minutes of the source passing through the beam of the LPA LPI at half power; 4) the maximum number of pulses registered per three minutes does not exceed four (J1132+25, J1336+33), and even two (J0640+07, J1005+30); 5) power law (J0640+07, J1132+25), and lognormal with a power law tail (J1005+30, J1336+33)  distribution of pulse amplitudes; 6) the excess of peak amplitudes and energies by tens and hundreds of times relative to the corresponding values for the average profiles, which is typical for pulsars with giant pulses.

The strongest change in activity was observed for J1336+33. For 524 days, it showed high activity when S/N was up to 41, and then for 1372 days, only solitary weak (at the detection level) pulses were observed with long time intervals between them. It is also possible that nulling is characterized by very weak emission and the sensitivity of the LPA LPI is not enough to detect it. Anyway, J1336+33 is significantly distinguished from all known RRATs by rapid changing of activity to long interval of (around three years) absence of emission or its strong attenuation. It would be interesting to study this object using timing to see if this rapid changing is caused by changing of period and period derivative. Unfortunately we did not have this possibility. Although the observation session lasts only three minutes, however, the appearance of single pulses is a random process and the absence of pulses over a long time interval during daily observation of sources could reflect a systematic and persistent change in the emission process.

The modulation index reflects amplitude variability in time and it is significantly different for J0640+07, J1005+30 ($m = 2.2$ and $1.25$) and 1132+25, 1336+33 ($m = 0.2$ and $0.5$). In this case, a small value of $m$ is caused mainly by long intervals of nulling or strong decreasing of emission. For RRATs  J0640+07, J1005+30 we have rare pulses but they are more smoothly distributed in time (see Fig. 1) and the ratio of  the amplitude of the strongest pulse to the mean amplitude $I_{max}/<I_2>$ for them is much larger than for another two pulsars. Both of these qualities collectively lead to $m > 1$.

A strong change in the character of emission can be associated with both external and internal causes. RRATs can be extinct pulsars that are re-activated due to the interaction of the pulsar's magnetosphere with the matter falling onto it, or the interaction of the magnetosphere with the surrounding matter (\citeauthor{Li2006}, \citeyear{Li2006}, \citeauthor{Luo2007}, \citeyear{Luo2007}). Internal causes can be determined by a sharp change in the primary plasma density or the conditions of emission propagation in the magnetosphere (\citeauthor{Timokhin2010}, \citeyear{Timokhin2010}).

The broken power law character of the distribution 
was previously obtained for RRAT J0139+33 (\citeauthor{Brylyakova2020}, \citeyear{Brylyakova2020}), which was also observed at the frequency of 111 MHz. This distribution pattern is one of the main features of giant pulses. In the paper of  \citeauthor{Karuppusamy2012} (\citeyear{Karuppusamy2012}) giant pulses are defined as pulses with an energy exceeding the energy of the average profile by more than 10 times. This ratio is quite arbitrary, and the RRATs for which the average profile was obtained (J1132+25, J1336+33) satisfy this condition. Studies at higher frequencies: 1400 and 820 MHz (\citeauthor{Cui2017}, \citeyear{Cui2017}, \citeauthor{Mickaliger2018}, \citeyear{Mickaliger2018}) had shown that for most RRATs the distribution of amplitudes and energies is mostly lognormal, as for ordinary pulsars, but a few RRATs had a log-normal distribution with a power law tail.

There is only one investigation with simultaneous observations RRAT at meter and decimeter wavelength (\citeauthor{Meyers2019}, \citeyear{Meyers2019}). They find that the single-pulse amplitude distribution is a truncated exponential at 150 MHz and a power low at 1400 MHz.

  Other studies of the pulse amplitude distribution in more than two dozen RRATs discovered in the decimeter wavelength range (\citeauthor{Cui2017}, \citeyear{Cui2017}, \citeauthor{Mickaliger2018}, \citeyear{Mickaliger2018}, \citeauthor{Shapiro-Albert2018}, \citeyear{Shapiro-Albert2018}) showed a lognormal or lognormal with power law tail distribution. \citeauthor{Meyers2019} (\citeyear{Meyers2019}) estimated the pulse energy distribution for J2325-0530 at 150 MHz and 1400 MHz but because of small sample of single pulses they were unable to drow concrete conclusions. Since our sample is small and does not have observations at higher frequences, it is impossible to conclude that the distribution of the pulse amplitude may depend on the frequency of observations for all RRATs. 
The analysis shows that RRATs are not a homogeneous sample, but a mixture of different kinds of pulsars, and long-term studies of large RRATs samples are needed, which will allow us to judge more accurately the validity of different hypotheses. 

\section{Data availability}

The row data underlying this article will be shared on reasonable request to the corresponding author. The tables with S/N of pulses are in http://prao.ru/online\%20data/onlinedata.html 

\section*{Acknowledgements}
The PYTHON LMFIT package\footnote{https://zenodo.org/record/4516651} was used for data analysis. The authors express their appreciation to V.M.Malofeev, V.A. Potapov, A.N.Kazantzev for useful discussions in the course of the work. Special thanks to Yu.Yu. Kovalev for a donated server that was used to process observations data.


\end{document}